\documentclass[final,5p,times,twocolumn]{elsarticle}

\usepackage{amssymb}
\usepackage{amsmath}
\usepackage{balance}

\usepackage[colorlinks,urlcolor=blue,citecolor=blue,linkcolor=blue]{hyperref}

\journal{Journal of Magnetism and Magnetic Materials}

\begin{document}


\begin{frontmatter}

\title{Effect of granules anisotropy on ``double~quantum'' magnetic resonance excitation in~nanogranular composites}

\author[IPP]{A.~B.~Drovosekov}
\ead{drovosekov@kapitza.ras.ru}

\author[IPP]{M.~Yu.~Dmitrieva}

\author[VSU,NRC]{A.~V.~Sitnikov}

\author[NRC]{S.~N.~Nikolaev}

\author[NRC,ITAE,IRE]{V.~V.~Rylkov}

\affiliation[IPP]{organization={P.L.Kapitza Institute for Physical Problems, RAS},
city={Moscow},
postcode={119334},
country={Russia}}

\affiliation[VSU]{organization={Voronezh State Technical University},
city={Voronezh},
postcode={394026},
country={Russia}}

\affiliation[NRC]{organization={National Research Centre Kurchatov Institute},
city={Moscow},
postcode={123182},
country={Russia}}

\affiliation[ITAE]{organization={Institute of Theoretical and Applied Electrodynamics, RAS},
city={Moscow},
postcode={125412},
country={Russia}}

\affiliation[IRE]{organization={Kotelnikov Institute of Radio Engineering and Electronics, Fryazino Branch, RAS},
city={Fryazino, Moscow region},
postcode={141190},
country={Russia}}

\begin{abstract}
Films of metal-insulator nanogranular composites (CoFeB)$_x$(Al$_2$O$_3$)$_{100-x}$ with different contents of the metal ferromagnetic (FM) phase CoFeB ($x\approx15{-}50$~at.\,\%) are investigated by the method of electron spin resonance (ESR) in a wide range of frequencies ($f=7{-}80$~GHz) and temperatures ($T=4.2{-}300$~K). Besides the conventional FM resonance signal, the experimental spectra demonstrate an~additional absorption peak with a double effective g-factor $g_\mathrm{eff}\approx4$ which is explained within the quantum mechanical ``giant spin'' model by excitation of ``double quantum'' transitions in FM granules CoFeB. According to the theory, the intensity of this ``double quantum'' peak is a complex function of frequency and temperature, including as parameters the granule magnetic moment and anisotropy. Experimentally, the size and anisotropy of the granules can be varied either changing the nominal FM phase content $x$ in the composites or annealing the samples at different temperatures. Here we study the effects of concentration $x$ and thermal annealing of (CoFeB)$_x$(Al$_2$O$_3$)$_{100-x}$ films on their ESR spectral parameters. The observed behavior of the ``double quantum'' peak intensity is well explained within the considered ``giant spin'' theoretical concept. In conclusion, we demonstrate the correlation between the size of FM granules in nanocomposites and their anisotropy, indicating the surface origin of this anisotropy.
\end{abstract}

\end{frontmatter}

\section{Introduction}

In many cases, properties of fine magnetic particles prove to be quite different from bulk magnetic materials. The role of surface effects and thermal fluctuations grows extremely for nanoscale objects. Moreover, manifestations of quantum phenomena are possible. Magnetization dynamics in superparamagnetic ensembles of single domain ferromagnetic (FM) particles demonstrates features that can be treated in terms of both classical and quantum approaches \cite{Atsarkin2020}. Electron spin resonance (ESR) provides a simple way to reveal such unusual dynamical effects in superparamagnetic systems \cite{Noginova2007, Noginova2008, Fittipaldi2016, Domracheva2017, Drov2024, Drov2026}.

In the works \cite{Drov2024, Drov2026}, we studied ESR in films of metal-insulator nanogranular composites M$_x$D$_{100-x}$ with different compositions and percentage ratio $x$ of metal and dielectric phases (``M'' and ``D''). The structures represent an array of metallic FM nanogranules randomly distributed in a solid-state insulating medium (matrix). Depending on the nominal FM phase content $x$, the magnetic moment of the granules $\mu$ varies in the range $\mu\sim10^2{-}10^4\mu_\text{B}$, where $\mu_\text{B}$ is Bohr magneton.

It was found that, besides the conventional signal of FM resonance (FMR), the experimental spectra contained an additional ``anomalous'' absorption peak demonstrating a number of unusual properties: double effective g-factor $g_\mathrm{eff}\approx4$, non-standard conditions for its excitation by longitudinal microwave magnetic field, and nonmonotonic dependence of the peak intensity on temperature.

It was shown that many of the observed features of ESR spectra can be well explained, if we consider FM nanogranules as quantum objects (paramagnetic centers) with ``giant'' spins $S \sim 10^2{-}10^4$ \cite{Atsarkin2020, Noginova2007, Noginova2008, Fittipaldi2016, Domracheva2017, Drov2024}. Within this approach, the appearance of a~resonance peak with effective g-factor $g_\mathrm{eff}\approx4$ is associated with the excitation of double quantum transitions in the granules with a change in the spin projection $\Delta m = \pm2$. An~equivalent classical interpretation suggests that the absorption of microwave power by classical magnetic moments occurs at multiple harmonics of their precessional motion in the presence of magnetic anisotropy or dipole-dipole interactions between the particles \cite{Drov2026}. From this point of view, the observed ``double quantum'' peak corresponds to excitation of second precessional harmonics in the particles.

According to the theory \cite{Noginova2008, Drov2026}, the intensity of such ``double quantum'' peak is a~complex function of frequency and temperature, including as parameters the granule magnetic moment, anisotropy, and interparticle dipolar interactions. Thus, the possibility to control these parameters experimentally becomes important for further verification of the considered model.

Previously, it was shown in many works that the size of the granules in nanocomposites can be varied either changing the nominal FM phase content or subjecting the samples to heat treatment \cite{Drov2024, Stognei2001, Kalinin2001, Stognei2003, Denisova2016, Review}. In this work, we perform systematic studies of ESR in (CoFeB)$_x$(Al$_2$O$_3$)$_{100-x}$ nanocomposite films with different FM phase contents $x$. We study the structures in their initial state after sputtering, as well as investigate the effects of thermal annealing of the samples. As a result, we analyze applicability of the ``giant spin'' concept to describe the experimental spectra, and establish mechanisms responsible for excitation of ``double quantum'' transitions in the granules.

\section{Samples and experimental techniques}

The studied nanogranular films (CoFeB)$_x$(Al$_2$O$_3$)$_{100-x}$ with a~thickness of about $2{-}3~\mu$m were synthesized on glass-ceramic substrates by ion-beam sputtering of composite targets \cite{Granovsky2016, Rylkov2017}. The target represents a flat plate of FM alloy Co$_{40}$Fe$_{40}$B$_{20}$ (CoFeB), on which a set of rectangular Al$_2$O$_3$ strips are placed. Uneven arrangement of the strips on the target surface results in a formation of nanocomposite film with a smooth gradient of the FM phase content $x$ along the substrate ($\sim1$~at.\,\%/cm). The resulting total change of the concentration $x$ along the substrate covers a wide range from $x\approx15$ to 60~at.\,\%. Further studies are carried out on small ($\approx5\times5$~mm) pieces of the sputtered film, so that the change of $x$ within one sample is negligible.

Detailed investigations of structure, electron transport, static magnetization, and magnetooptical properties of the synthesized CoFeB--Al$_2$O$_3$ composites were performed in a set of previous works \cite{Review, Rylkov2017, Rylkov2018, Ganshina2020}.

According to the transmission electron microscopy, the obtained structures consist of FM nanogranules CoFeB randomly distributed inside the amorphous insulating Al$_2$O$_3$ matrix \cite{Rylkov2017, Rylkov2018}. The granules have approximately spherical shape and their average size (2--5~nm) increases smoothly with a growth of the FM phase content $x$ in the nanocomposite.

Electron transport measurements showed that the percolation threshold in the films corresponds to FM phase content of about $x_\text{p}\approx57$~at.\,\%, while the metal-insulator transition takes place at slightly lower concentration $x_\text{c}\approx48$~at.\,\% \citep{Rylkov2017}. According to static magnetization and magnetooptical data, a crossover from superparamagnetic to ferromagnetic behavior of the samples occurs approximately in the same concentration range $x\approx50{-}55$~at.\,\% \cite{Review, Rylkov2018, Ganshina2020}.

In this work, a set of (CoFeB)$_x$(Al$_2$O$_3$)$_{100-x}$ films with $x\approx15{-}50$~at.\,\% is studied by electron magnetic (spin) resonance. Room-temperature spectra are measured in a wide range of frequencies $f=7{-}80$~GHz. Temperature dependence is investigated at a fixed frequency $f\approx25$~GHz in the range $T=4.2{-}300$~K. The experimental spectra are recorded in field-sweep regime in a magnetic field (up to~24~kOe) applied in the film plane. At frequencies 7--38~GHz, a cavity-based transmission-type spectrometer is used. At higher frequencies up to 80~GHz, a reflection from shorted rectangular waveguide with the studied sample inside is measured.

To study the effect of thermal treatment, several pieces of nanocomposite samples (CoFeB)$_{33}$(Al$_2$O$_3$)$_{67}$ were annealed for 15~min.\ in muffle furnace at different temperatures $T_\text{ann}$ in the interval $T_\text{ann}=500{-}800^\circ$C.

\section{Theoretical background}\label{Theory}

\subsection{FMR line position in superparamagnetic film}

According to the well known Kittel formula, the FMR frequency for a thin magnetic film placed in tangential external field $H$ is determined by expression
\begin{equation}\label{KittelFreq}
\omega = \gamma \sqrt{H(H+4\pi M)},
\end{equation}
where $\gamma$ is gyromagnetic ratio and $M$ is sample magnetization. Using this equation, the resonance field $H_\text{res}$ measured at constant frequency can be easily calculated:
\begin{equation}\label{KittelField}
H_\text{res} = \sqrt{H_0^2 + (2\pi M)^2} - 2\pi M,
\end{equation}
where we introduce for convenience
\begin{equation*}
H_0 = \frac{\omega}{\gamma} = \frac{f}{\gamma/2\pi}
\end{equation*}
($H_\text{res} = H_0$, when $M\rightarrow0$). If the film is in superparamagnetic state, we must take into account the dependence of the magnetization $M$ on the external field and temperature $T$:
\begin{equation}\label{MvsHT}
M = M_\text{S} L(\varkappa_1),
\end{equation}
where $M_\text{S}$ is saturation magnetization, $L(x) = \tanh^{-1}x - x^{-1}$ is~Langevin function, and
\begin{equation}\label{kappa}
\varkappa_1 = \frac{\mu (H_\text{res} + 4\pi M/3)}{k_\text{B}T},
\end{equation}
where $\mu$ is granule magnetic moment and $k_\text{B}$ is Boltzmann constant. Here we bare in mind the presence of the Lorentz field $4\pi M/3$ inside the film in addition to the external filed \cite{Drov2026, Godinho1995, Dubowik1996}.

Solving the equation system \eqref{KittelField}--\eqref{kappa}, we can obtain $H_\text{res}$ as a function of temperature $T$ and frequency $f$ for given material parameters $M_\text{S}$ and $\mu$. Strictly speaking, Eq.~\eqref{MvsHT} is valid only above the blocking temperature $T_\text{B}$ which, however, is estimated not higher than $T_\text{B}\sim20$~K, if the films are not too close to the percolation threshold ($x\lesssim50$~at.\,\%)\footnote{The blocking temperature is defined by relation $T_\text{B}\approx\mu H_\text{A}/25k_\text{B}$, where $H_\text{A}$ is an effective field of granules anisotropy. For typical experimental values $\mu\approx4\cdot10^3\mu_\text{B}$ and $H_\text{A}\approx2$~kOe, we estimate $T_\text{B}\approx20$~K.}.

We also note that Eqs.~\eqref{KittelFreq},~\eqref{KittelField} are applicable when the FMR linewidth $\Delta H$ is sufficiently small $\Delta H \ll H_\text{res}$. Otherwise, the effects of inhomogeneous line broadening may lead to additional shift of the resonance peak position \cite{Raikher1992, Gurevich}. In practice, the relation $\Delta H \ll H_\text{res}$ proves to be correct for most investigated nanocomposite samples except the case of very low FM phase contents $x\lesssim30$~at.\,\% and temperatures $T\lesssim50$~K.

\subsection{Intensities of FMR and ``double quantum'' peaks}

The integral intensity $I_\text{1Q}$ of the main (``single quantum'') resonance line in (super)paramagnetic medium is known to be propotional to the sample magnetization:
\begin{equation*}
I_\text{1Q} \propto M \propto L(\varkappa_1).
\end{equation*}
However, if the sample shape is different from spherical and the magnetization $M$ is large or comparable with the external field, this expression requires correction due to essential demagnetizing effects \cite{Gurevich}.

In our experimental geometry, when the static and microwave magnetic fields are applied in the plane of thin film, the integral intensity of the FMR peak is proportional to \cite{Gurevich}
\begin{equation}\label{IntFMR}
I_\text{FMR} \propto \frac{H_\text{res} + 4\pi M}{H_\text{res} + 2\pi M} L(\varkappa_1),
\end{equation}
where $H_\text{res}$, $M$, and $\varkappa_1$ are defined by the system \eqref{KittelField}--\eqref{kappa}.

As it was shown previously \cite{Noginova2008, Drov2026}, the ``double quantum'' resonance peak may arise in an ensemble of superparamagnetic particles due to the presence of magnetic anisotropy in the granules or dipolar interactions between them. In both cases, the integral intensity of the ``double quantum'' peak $I_\text{2Q}$ can be calculated within the perturbation theory in the limit of high fields.

In particular, if weakly anisotropic particles are considered with an effective uniaxial anisotropy field $H_\text{A} \ll H_0$, the intensity $I_\text{2Q}$ is proportional to
\begin{equation}\label{Int2Q}
I_\text{2Q} \propto \frac{8}{15} \left( \frac{H_\text{A}}{H_0} \right)^2 \frac{\varkappa - 3L(\varkappa)}{\varkappa^2},
\end{equation}
where
\begin{equation*}
\varkappa = \frac{\mu H_0}{2 k_\text{B} T}.
\end{equation*}
In the presence of weak dipole-dipole interactions between the particles, $I_\text{2Q}$ is propotional to
\begin{equation}\label{Int2Qdip}
I_\text{2Q}^\text{dip} \propto \frac{2}{15} \left( \frac{H_\text{D}}{H_0} \right)^2 \frac{L^2(\varkappa)}{\varkappa},
\end{equation}
where $H_D = 3\mu/r^3$ is effective dipolar field at the distance $r$ between the particles ($H_\text{D} \ll H_0$).

Eqs.~\eqref{Int2Q} and \eqref{Int2Qdip} have qualitatively similar dependence on~$\varkappa$. Both functions are non-monotonic on temperature. In the limit of low temperatures ($\varkappa\gg1$), they demonstrate linear decrease $I_\text{2Q}\propto T$. At~high temperatures ($\varkappa\ll1$), they follow the Curie law $I_\text{2Q}\propto T^{-1}$. The maximal value $I_\text{2Q}$ is achieved at $T_\mathrm{max} \approx 0.3\,\mu H_0/k_B$ which is dependent on the particle magnetic moment $\mu$.

Note that in Eqs.~\eqref{IntFMR}--\eqref{Int2Qdip}, we omit the proportionality coefficient which depends on experimental setup parameters. In~practice, the difficulty to achieve accurate calibration of this coefficient prevents extraction of the absolute values $H_\text{A,D}$ from experimental dependence $I_\text{2Q}(T)$. On the contrary, measuring the relation $I_\text{2Q}/I_\text{FMR}$ removes this uncertainty and allows direct determination of the parameters $H_\text{A,D}$.

The temperature dependence of the relation $I_\text{2Q}/I_\text{FMR}$ is to some extent similar to $I_\text{2Q}(T)$ \cite{Noginova2008}. However, the maximum of this dependence is not so pronounced, and at high temperatures ($\varkappa\ll1$), the relation $I_\text{2Q}/I_\text{FMR}$ saturates to constant value:
\begin{equation}\label{Simple}
\frac{I_\text{2Q}}{I_\text{FMR}} = \frac{4}{75} \left( \frac{H_\text{A}}{H_0} \right)^2,  \qquad  \frac{I_\text{2Q}^\text{dip}}{I_\text{FMR}} = \frac{1}{45} \left( \frac{H_\text{D}}{H_0} \right)^2,
\end{equation}
depending on the experimental frequency as $H_0^{-2}\propto f^{-2}$. Note, that these two formulas are equivalent when $H_\text{D}\approx1.55H_\text{A}$.

\begin{figure}[t]
\centering
\includegraphics[width=.85\columnwidth]{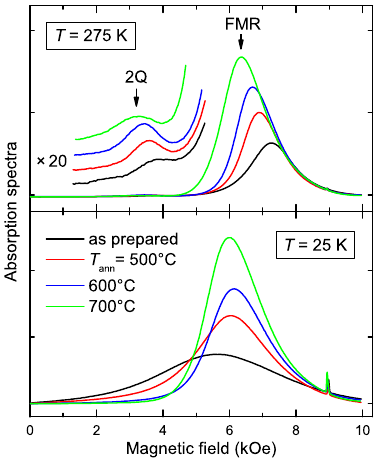}
\caption{Magnetic resonance spectra measured at $T=275$ and 25~K for nanocomposite films (CoFeB)$_{33}$(Al$_2$O$_3$)$_{67}$ annealed for 15~min.\ at different temperatures $T_\text{ann}$. The spectra are obtained at frequency $f\approx25.0$~GHz in magnetic field applied in the film plane.}
\label{Fig1}
\end{figure}

\section{Results and discussions}

\subsection{Frequency dependence of ESR spectral parameters}

\begin{figure}[t]
\centering
\includegraphics[width=0.91\columnwidth]{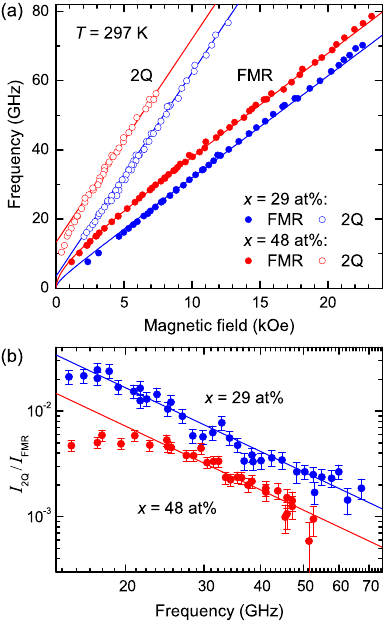}
\caption{(a) Room-temperature frequency-field dependence $f(H)$ for FMR and ``double quantum'' (2Q) peaks in (CoFeB)$_x$(Al$_2$O$_3$)$_{100-x}$ films with different FM phase contents $x$. (b)~The relation of the absorption peak intensities $I_\text{2Q}/I_\text{FMR}$ as a function of frequency (double logarithmic scale). Symbols are experimental data, lines are calculations (see text).}
\label{Fig2}
\end{figure}

Magnetic resonance spectra of the investigated nanogranular films demonstrate the presence of a pronounced FMR absorption peak and an additional weak ``double quantum'' (2Q) peak located at about half field with respect to FMR (Fig.~\ref{Fig1}). According to the Kittel formula \eqref{KittelFreq}, an increase of the film magnetization leads to a shift of the FMR peak to weaker fields. The position of the ``double quantum'' peak approximately corresponds to double FMR frequency, however, more accurately, its frequency is defined by equation \cite{Drov2026}
\begin{equation}\label{Freq2Q}
\omega = 2\gamma \left( H + 4\pi M/3 \right).
\end{equation}
Fig.~\ref{Fig2}a demonstrates frequency-field diagrams for the absorption lines in two films with different FM phase contents~$x$. Experimental data are well fitted by Eqs.~\eqref{KittelFreq} and \eqref{Freq2Q} for FMR and ``double quantum'' peaks respectively. For both samples, the gyromagnetic ratio $\gamma/2\pi\approx2.94$~GHz/kOe corresponds to the effective $g$-factor $g_\mathrm{eff}\approx2.1$, typical of CoFeB alloys\footnote{Further we use $g$-factor value $g_\mathrm{eff}=2.1$ as a fixed parameter to fit the experimental data.}. The magnetization increases from $4\pi M\approx1.8$~kOe at $x\approx29$~at.\,\% to $4\pi M\approx6.8$~kOe at $x\approx48$~at.\,\%.

Fig.~\ref{Fig2}b shows the relation of absorption peak intensities $I_\text{2Q}/I_\text{FMR}$ as a function of frequency for the considered two films. The experimental dependences in the figure are well fitted by simplified Eqs.~\eqref{Simple} in the region of high frequencies. The discrepancy at low frequencies can be due to inapplicability of the perturbation theory used to derive the equations. The estimated effective field parameters are $H_\text{A}\approx3.8$~kOe (or $H_\text{D}\approx5.9$~kOe) for $x\approx29$~at.\,\% and $H_\text{A}\approx2.5$~kOe (or $H_\text{D}\approx3.9$~kOe) for $x\approx48$~at.\,\%.

Obviously, the dipolar interactions $H_\text{D}$ in the films should diminish with decreasing FM phase content, while the experiment shows inverse dependence. Thus, we consider the anisotropy of granules as the main mechanism initiating ``double quantum'' excitations in nanocomposite films.

Note that the magnetocrystalline anisotropy of bulk CoFeB alloys is estimated to be not higher than $\sim100$~Oe \cite{Sundar2005} which is much smaller than the obtained values $H_\text{A}\sim1$~kOe. Thus, it is natural to suppose that this anisotropy has surface origin, and arises near the interface between the metallic granules and oxide matrix. Relative contribution of the surface energy to the total magnetic energy is growing as the granules become smaller. This explains qualitatively the observed increase in $H_\text{A}$ with the decrease of the FM phase content in the nanocomposite.

For example, within the simplest model, magnetic moment~$\mu$ of a spherical granule is proportional to its volume $V$: $\mu = M_0 V$, where $M_0$ is saturation magnetization of bulk CoFeB alloy. The anisotropy energy $E_\textbf{A}$ is defined by the granule surface $S$ via relation $E_\textbf{A} = K_\text{S}S$, where $K_\text{S}$ is anisotropy constant. Taking into account that $H_\text{A} = 2E_\textbf{A}/\mu$ and $S = (36\pi)^{1/3} V^{2/3}$, we get
\begin{equation}\label{Ha}
H_\text{A} = \frac{2K_\text{S}}{M_0} \left( \frac{36\pi M_0}{\mu} \right)^{1/3}.
\end{equation}
Thus, we can expect the anisotropy of granules $H_\text{A}$ to reduce with their magnetic moment as $H_\text{A} \propto \mu^{-1/3}$.

The magnetic moment $\mu$ of the granules can be determined from temperature dependence of the FMR line position $H_\text{res}(T)$ and the ``double quantum'' peak intensity $I_\text{2Q}(T)$, as it is shown in Section~\ref{Theory}. At~the same time, the anisotropy of granules can be estimated from experimental relation $I_\text{2Q}/I_\text{FMR}(T)$. Below we perform detailed analysis of such dependences for films with different FM phase contents and films annealed at different temperatures to study the applicability of the considered theoretical concepts and to establish the correlation between the granule size and anisotropy.

\subsection{Temperature dependence of ESR spectral parameters}

\subsubsection{Nanocomposites with different FM phase contents}

\begin{figure}[t]
\centering
\includegraphics[width=.9\columnwidth]{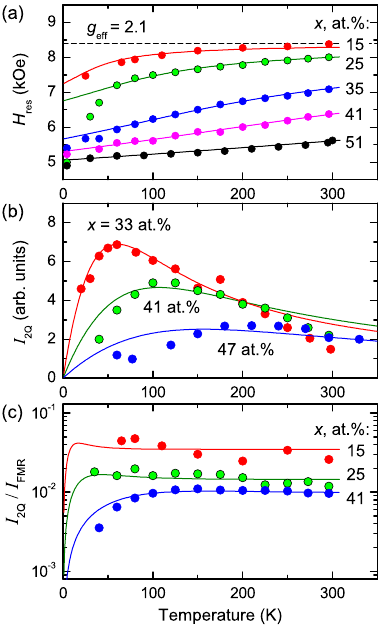}
\caption{Temperature dependence of the FMR line position $H_\text{res}$ (a), the ``double quantum'' peak intensity $I_\text{2Q}$ (b), and the relation $I_\text{2Q}/I_\text{FMR}$ (c) obtained at frequency $f\approx24.7$~GHz for nanocomposite films (CoFeB)$_x$(Al$_2$O$_3$)$_{100-x}$ with different FM phase contents~$x$. Symbols are experimental data, solid lines are their approximation (see text). The dashed line in the plot (a) corresponds to $H_\text{res}=H_0$ for the effective $g$-factor $g_\mathrm{eff}=2.1$.}
\label{Fig3}
\end{figure}

Fig.~\ref{Fig3} demonstrates experimental dependences $H_\text{res}(T)$, $I_\text{2Q}(T)$, and $I_\text{2Q}/I_\text{FMR}(T)$ for nanocomposite films with different FM phase contents. For every sample, the experimental data are well approximated by theoretical curves calculated from the equation system \eqref{KittelField}--\eqref{Int2Q}, using only three fitting parameters: $M_\text{S}$, $\mu$, and $H_\text{A}$.

The experimental dependences $H_\text{res}(T)$ demonstrate monotonic increase. According to the theoretical model described in Section~\ref{Theory}, the values $H_\text{res}(0)$ extrapolated to $T=0$ define the saturation magnetization $M_\text{S}$, while the slope of the curves $H_\text{res}(T)$ is determined by granules magnetic moment $\mu$. The dependences $I_\text{2Q}(T)$ show characteristic maximums at $T_\mathrm{max} \approx 0.3\,\mu H_0/k_B$. The ratios $I_\text{2Q}/I_\text{FMR}(T)$ at high temperatures reach constant values defined by granules effective anisotropy field $H_\text{A}$ in agreement with Eq.~\eqref{Simple}.

\begin{figure}[t]
\centering
\includegraphics[width=0.9\columnwidth]{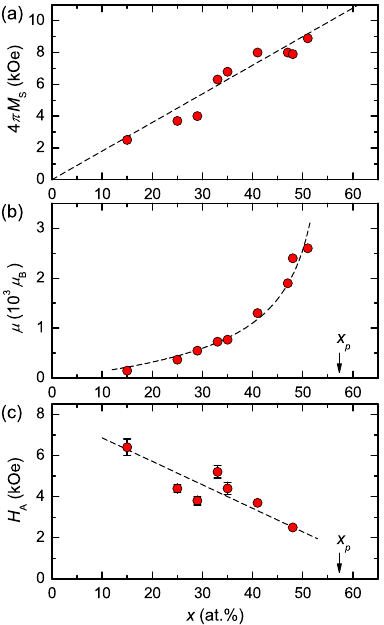}
\caption{The saturation magnetization $M_\text{S}$ (a), the magnetic moment of granules $\mu$ (b), and their effective anisotropy field $H_\text{A}$ (c) as a function of the FM phase content~$x$ in nanocomposite films (CoFeB)$_x$(Al$_2$O$_3$)$_{100-x}$. Symbols are obtained from fitting the resonance peak parameters (see Figs.~\ref{Fig2} and~\ref{Fig3}), lines are guides for an eye. The arrow labeled $x_\text{p}$ indicates the percolation thresold according to \cite{Rylkov2017}.}
\label{Fig4}
\end{figure}

In Fig.~\ref{Fig4}, the resulting fitting parameters $M_\text{S}$, $\mu$, $H_\text{A}$ are plotted as a~function of the FM phase content $x$ in the films. It is interesting to note that the experimental dependence $4\pi M_\text{S} (x)$ is close to linear. Its formal extrapolation to $x=100$~at.\,\% gives $4\pi M_0 \approx 18$~kOe ($M_0 \approx 1400$~emu/cm$^3$) which is typical of bulk CoFeB alloys \cite{Liu2011}. The granule magnetic moment $\mu$ also grows with $x$, however, the dependence $\mu(x)$ is not linear, increasing sharply as $x$ approaches the percolation threshold $x_\text{p}\approx57$~at.\,\%. This behavior seems natural as we expect the granules to form macroscopic percolation clusters at $x = x_\text{p}$.

The granule anisotropy parameter $H_\text{A}$ demonstrates a tendency to decrease with $x$ (Fig.~\ref{Fig4}c). The lowest concentration $x\approx15$~at.\,\% corresponds to smallest granules with $\mu\approx150\mu_\text{B}$ and highest $H_\text{A} \approx 6.4$~kOe. These results confirm the suggestion that the observed ``double quantum'' peak originates from the presence of large surface magnetic anisotropy of the nanogranules.

\subsubsection{Effect of samples annealing}

Another way to vary the size of FM granules in metal-insulator nanocomposites is thermal annealing of the samples \cite{Stognei2003, Denisova2016, Review}. In our experiments, nanocomposite films CoFeB--Al$_2$O$_3$ with the same content of the FM phase $x\approx33$~at.\,\% were annealed for 15~min.\ at different temperatures $T_\text{ann}$ in the interval $T_\text{ann}=500{-}800^\circ$C. As we found, annealing the samples at lower temperatures does not lead to noticeable change of the ESR spectra. On the contrary, increasing $T_\text{ann}$ from 500$^\circ$C to 800$^\circ$C initiates essential modification of ESR spectra (Fig.~\ref{Fig1}).

The room-temperature spectra show that the main FMR peak of the annealed samples is shifting to lower fields and becomes more intense, indicating the increase of the film magnetization. This increase can be explained by the growth of the granules and weakening thermal fluctuations of their magnetic moments. On the contrary, the position of the FMR peak at low temperature is almost independent on $T_\text{ann}$. This behavior reflects the fact that the annealing procedure does not affect the saturation magnetization of the films. Indeed, the atomic diffusion processes during thermal annealing are not expected to change the total amount of the FM phase in the samples. At the same time, the annealed nanocomposites demonstrate an essential narrowing of the FMR line at low temperatures. Qualitatively, this behavior indicates the suppression of inhomogeneities in the films, and can be explained by the reduction of granules anisotropy \cite{Raikher1992, Gurevich}.

\begin{figure}[t]
\centering
\includegraphics[width=0.9\columnwidth]{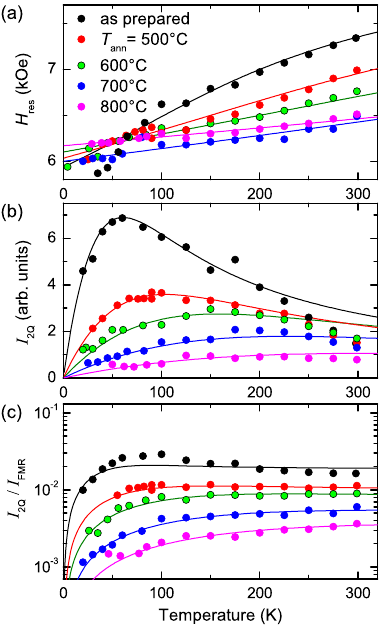}
\caption{Temperature dependence of the FMR line position $H_\text{res}$ (a), the ``double quantum'' peak intensity $I_\text{2Q}$ (b), and the relation $I_\text{2Q}/I_\text{FMR}$ (c) obtained at frequency $f\approx25.0$~GHz for nanocomposite films (CoFeB)$_{33}$(Al$_2$O$_3$)$_{67}$ annealed for 15~min.\ at different temperatures $T_\text{ann}$. Symbols are experimental data, lines are their approximation (see text).}
\label{Fig5}
\end{figure}

Fig.~\ref{Fig5} demonstrates the resulting experimental dependences $H_\text{res}(T)$, $I_\text{2Q}(T)$, and $I_\text{2Q}/I_\text{FMR}(T)$ for films annealed at different temperatures. Qualitatively, Fig.~\ref{Fig5} resembles Fig.~\ref{Fig3} since the effect of annealing is, to some extent, similar to the effect of increasing FM phase concentration, both leading to the growth of the granules dimensions. The experimental data in Fig.~\ref{Fig5} are also well fitted by formalism described in Section~\ref{Theory}, using Eqs.~\eqref{KittelField}--\eqref{Int2Q}.

In agreement with the qualitative analysis, the saturation magnetization proves to remain almost unchanged after annealing the films at different temperatures: $4\pi M_\text{S} \approx 6.0\pm0.2$~kOe. At the same time, the gradual increase of $T_\text{ann}$ up to 800$^\circ$C leads to a significant growth of the granules magnetic moments $\mu$ (from about $1000$ to $4000\mu_\text{B}$) with simultaneous decrease of the effective anisotropy field $H_\text{A}$ (from about 5 down to 2~kOe).

\begin{figure}[t]
\centering
\includegraphics[width=0.85\columnwidth]{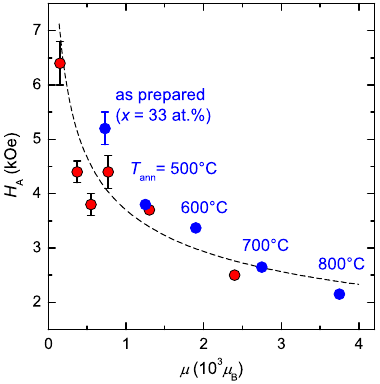}
\caption{Effective anisotropy field $H_\text{A}$ plotted as a function of the granule magnetic moment $\mu$ for nanocomposite films (CoFeB)$_x$(Al$_2$O$_3$)$_{100-x}$ with different FM phase contents~$x$ (red circles) and (CoFeB)$_{33}$(Al$_2$O$_3$)$_{67}$ annealed at different temperatures $T_\text{ann}$ (blue circles). The dashed line corresponds to $H_\text{A} \propto \mu^{-1/3}$ as~predicted by Eq.~\eqref{Ha}.}
\label{Fig6}
\end{figure}

The experimental parameters $\mu$ and $H_\text{A}$ for different studied samples are summarized in $\mu{-}H_\text{A}$ diagram in Fig.~\ref{Fig6}, where the effects of both FM phase content and thermal annealing are presented. As it is shown in figure, the experimental dependence $H_\text{A}(\mu)$ approximately follows the power low $H_\text{A} \propto \mu^{-1/3}$ predicted by Eq.~\eqref{Ha}. From this fitting, the constant $K_\text{S}$ of the surface anisotropy is estimated as $K_\text{S}\approx0.1$~erg/cm$^2$.

\section{Conclusion}

Using the method of electron spin resonance, we studied films of metal-insulator nanogranular composites CoFeB--Al$_2$O$_3$ representing ensembles of metallic superparamagnetic particles CoFeB randomly distributed in insulating Al$_2$O$_3$ oxide matrix. The experimental spectra, besides the conventional FMR signal, demonstrate the presence of additional ``anomalous'' absorption peak with a double effective g-factor $g_\mathrm{eff}\approx4$. Within the quantum mechanical ``giant spin'' model, this peak is attributed to excitation of double quantum transitions in ferromagnetic nanogranules with a change in the spin projection $\Delta m = \pm2$. According to the model, the possibility of such excitations in an ensemble of superparamagnetic particles may arise due to the presence of magnetic anisotropy in the granules or dipolar interactions between them.

Here we have shown that the experimental features of the magnetic resonance spectra are well explained within the proposed theoretical concepts. Specific frequency and temperature dependences of the ``double quantum'' resonance peak intensity demonstrate a good agreement with the predictions of the ``giant spin'' model.

Studying composites with various concentrations of the FM phase and annealing the samples at different temperatures, we found a systematic correlation between the size of the granules and the observed intensity of the ``double quantum'' peak. The obtained results indicate that the presence of surface magnetic anisotropy in the granules is the main mechanism responsible for manifestation of ``double quantum'' excitations in the resonance spectra.

\section*{Acknowledgements}

The work was carried out within the framework of a state assignment and was financially supported by the Russian Science Foundation (project 22-12-00259-$\Pi$). The authors thank Andrey~B.~Drovosekov for assistance in annealing the samples.

\balance

\biboptions{sort&compress}

\end{document}